\documentclass[twocolumn,showpacs,preprintnumbers,amsmath,amssymb]{revtex4}

\usepackage{graphicx}
\usepackage{dcolumn}
\usepackage{bm}

\begin{document}


\title{Statistical-mechanical iterative algorithms on complex networks}

\author{Jun Ohkubo}
 \email[Email address: ]{jun@smapip.is.tohoku.ac.jp}
\author{Muneki Yasuda}
\affiliation{
Department of System Information Sciences, Graduate School of Information Sciences,
Tohoku University, 6-3-09, Aramaki-Aza-Aoba, Aoba-ku, Sendai 980-8579, Japan
}

\author{Kazuyuki Tanaka}
 \email[Email address: ]{kazu@smapip.is.tohoku.ac.jp}
\affiliation{
Department of Applied Information Sciences, Graduate School of Information Sciences,
Tohoku University, 6-3-09, Aramaki-Aza-Aoba, Aoba-ku, Sendai 980-8579, Japan
}


\begin{abstract}
The Ising models have been applied for various problems
on information sciences, social sciences, and so on.
In many cases, solving these problems corresponds to minimizing the Bethe free energy.
To minimize the Bethe free energy, a statistical-mechanical iterative algorithm is often used.
We study the statistical-mechanical iterative algorithm on complex networks.
To investigate effects of heterogeneous structures on the iterative algorithm,
we introduce an iterative algorithm based on information of heterogeneity of complex networks,
in which higher-degree nodes are likely to be updated more frequently than lower-degree ones.
Numerical experiments clarified that 
the usage of the information of heterogeneity affects the algorithm in Barab{\'a}si and Albert networks,
but does not influence that in Erd{\" o}s and R{\'e}nyi networks.
It is revealed that information of the whole system propagates rapidly
through such high-degree nodes in the case of Barab{\'a}si-Albert's scale-free networks.
\end{abstract}


\maketitle

\section{INTRODUCTION}

In recent years, complex networks have been studied a lot 
because of their importance as backbones of real complex systems \cite{Albert2002,Dorogovtsev2003}.
Many models have been proposed to capture properties of real-world networks;
the representatives of them are small-world networks proposed by Watts and Strogatz \cite{Watts1998},
and scale-free networks proposed by Barab{\'a}si and Albert \cite{Barabasi1999a}.
Much elaboration has made clear their topological features
such as the small-world property, the clustering property, and community structures.
Not only their topological features, but also their dynamical properties have been clarified.
For example, it has been revealed that 
the scale-free structure has large influence on dynamics in its heterogeneous structure
\cite{Satorras2001,Madar2004,Hayashi2004}.
It has been shown that the epidemic threshold vanishes
if their degree distributions have a power-law form, $P(k) \sim k^{-\gamma}$,
where the characteristic degree exponent $\gamma$ is lower than $3$ \cite{Satorras2001}.
Other dynamics are also important in order to understand characteristic properties of heterogeneous structures.
Indeed, there are many examples of dynamics on complex networks,
in which nodes are interacting each other and a macroscopically functional behavior emerges.
Examples of these networks with dynamics are
neural networks in a brain \cite{swmed},
Hopfield models in complex networks \cite{Stauffer2003,McGraw2003,Kim2004a},
protein folding \cite{Kussell2002},
and combinatoric optimization problems \cite{Mezard2002}.

While the heterogeneous structures of the complex networks have large influence on their dynamics,
one can use efficiently the heterogeneous property to control their dynamical properties.
The usage of the heterogeneous properties has been studied a lot mainly 
from the viewpoint of statistical physics.
For instance, in epidemic dynamics,
targeted immunization schemes based on the degree hierarchy work efficiently, which recover the epidemic threshold
in a case of the degree distribution $P(k) \sim k^{-\gamma}$, where $\gamma \leq 3$ \cite{Satorras2002}.
Therefore, one expects that high-degree nodes, so-called \textit{hubs},
play important roles in the dynamics on the heterogeneous networks,
and hence, one could use the heterogeneity to make efficient algorithms in a wide variety of problems.
This is true in some cases, and an efficient local search algorithm has been proposed,
which efficiently uses information of high-degree nodes (hubs) \cite{Adamic2001,Kim2002,Moura2003}.

The Ising spin system is one of the simple dynamical models,
and has attracted a lot of attentions by their theoretical interests.
In recent years, their theoretical analyses on complex networks have been performed
\cite{Aleksiejuk2002,Dorogovtsev2002,Leone2002,Dorogovtsev2004a}.
The Ising spin system could represent various real world phenomena;
simple opinion dynamics in a society is described by two Ising spin states.
Moreover, the Ising spin system has been used for many problems on information sciences
such as an artificial intelligence, image processing, error-correcting codes, and optimization problems
\cite{Nishimori2001}.
These problems can be solved by minimizing the free energy with suitably selected Hamiltonian.
Though it is difficult to obtain exact solutions of these problems,
there are many cases in which approximate solutions by means of the Bethe approximation are sufficient 
to solve these problems.
A statistical-mechanical iterative algorithm, 
so-called the Belief Propagation method \cite{Pearl1988,Tanaka2002,Tanaka2003},
enables us to solve these problems.
The iterative algorithm yields solutions approximately equivalent to those by means of the Bethe approximation.
It has been clear that the iterative algorithm is useful
to solve many problems on information sciences \cite{Tanaka2002}.

Recently, Mooij and Kappen have studied the validity of the Bethe approximation \cite{Mooij2005,Mooij2005a}.
They have concluded that the Bethe approximation is more powerful than the naive mean field approximation.
A relationship between the convergence property of the iterative algorithm and the spin-glass phase transition
has also been discussed,
and it has been revealed that the iterative algorithm does not work enough well in the spin-glass phase.
However, effects of heterogeneity of complex networks on the iterative algorithm have not been discussed yet.

In the present paper, we discuss effects of heterogeneity of complex networks on the iterative algorithm.
To investigate the effects, we introduce a statistical-mechanical iterative algorithm
in which higher-degree nodes are likely to be updated more frequently than lower-degree ones.
Applying the newly proposed iterative algorithms to ordinary random networks and scale-free networks,
we reveal that information of the whole systems propagates rapidly through the high-degree nodes;
the effects enable the iterative algorithm to work efficiently.

The outline of the present paper is as follows.
In Sec.~II, we introduce an Ising spin system.
The Bethe free energy to be minimized is also shown in Sec.~II.
The statistical-mechanical iterative algorithms are explained in Sec.~III.
We show numerical results in Sec.~IV.
The numerical results show that the influence of heterogeneity appears remarkably 
in the case of the scale-free networks.
Finally, we draw the main conclusions in Sec.~V.

\section{THE ISING SPIN SYSTEM}
We explain an Ising spin model used in the present paper.
Mooij and Kappen \cite{Mooij2005,Mooij2005a} have discussed the spin-glass phase transitions using the similar model.

We describe a graph $G = (\Omega,E)$ as an undirected graph without closed one-edge loops and multiple edges,
where $\Omega = \{ 1, 2, \dots, N\}$ is a set of nodes
and $E \subseteq \left\{ (i,j) \, | \, 1 \leq i < j \leq N  \right\}$ is a set of edges.
We denote a set of neighbors on node $i$ as $\partial i$.
Because there are no closed one-edge loops and multiple edges in the graph $G$,
the adjacency matrix ${\bf A} = \{A_{ij} | \, i,j \in \Omega\}$ is defined as follows:
$A_{ij} = A_{ji} = 1$ if $(i,j) \in E$, and $A_{ij} = A_{ji} = 0$ otherwise.
The degree of node $i$, $k_i$, represents that node $i$ is connected to $k_i$ other nodes;
hence $k_i = \lvert \partial i \rvert = \sum_{j\in \Omega} A_{ij}$.

On the graph $G$, we consider a random Ising model which is described by the Hamiltonian 
\begin{align}
H = - \sum_{(i,j) \in E} J_{ij} s_i s_j,
\label{eq_H}
\end{align}
where $s_i$ represents an Ising spin on node $i$ and takes the value $+1$ or $-1$,
and $J_{ij}$ is an interaction between nodes $i$ and $j$.
Each interaction $J_{ij}$ is different from each other.
We here denote the probability density function of $J_{ij}$ as $P_J (J_{ij})$.
From the Hamiltonian of Eq.~\eqref{eq_H},
we calculate the corresponding Boltzmann distribution over the configurations,
${\bf s} = (s_1, s_2, \dots, s_N) \in \{-1, +1\}^N$, 
as follows:
\begin{align}
\mathbb{P}({\bf s})
&= \frac{1}{Z} e^{-\beta H} 
= \frac{1}{Z} \prod_{(i,j) \in E} \phi_{{ij}} (s_i, s_j), 
\end{align}
where $Z$ is the partition function (or one may say that $Z$ is the normalization constant),
and $\beta$ the inverse temperature.
The notation $\phi_{ij}(s_i,s_j) = \exp\left( \beta J_{ij} s_i s_j \right)$ is introduced for simplicity.

Next, we write the free energy of the Hamiltonian of Eq.~\eqref{eq_H} by means of the Bethe approximation.
To obtain the Bethe free energy, the cluster variation method is used,
in which the Bethe free energy is represented
as a function of the marginal probability distributions,
i.e., $p_{ij}(s_i,s_j)$ and $p_i(s_i)$.
As a result,
the Bethe free energy is written as follows [25,28]:
\begin{align}
\beta \, \mathcal{F}_\textrm{Bethe}&(\{p_i(s_i)\},\{p_{ij}(s_i,s_j)\}) \notag \\
&\equiv -  \sum_{(i,j)\in E} \sum_{s_i = \pm 1} \sum_{s_j = \pm 1} p_{ij}(s_i,s_j) \ln \phi_{ij}(s_i,s_j)  \notag \\
&\quad+ \sum_{i=1}^{N}(1-k_i) \sum_{s_i = \pm 1} 
p_i(s_i) \ln p_i (s_i) \notag \\
&\quad+ \sum_{(i,j) \in E} \sum_{s_i = \pm 1} \sum_{s_j = \pm 1}
p_{ij}( s_i, s_j ) \ln p_{ij} ( s_i, s_j ).
\label{eq_bethe}
\end{align}
In many problems on information sciences,
we require the solution minimizing the Bethe free energy of Eq.~\eqref{eq_bethe}.
We can easily see that the \textit{paramagnetic} solution is a possible solution:
\begin{align}
\langle s_i \rangle &\simeq \sum_{s_i = \pm 1} s_i p_i ( s_i ) = 0, \\
\langle s_i s_j \rangle &\simeq \sum_{s_i = \pm 1} \sum_{s_j = \pm 1} s_i s_j p_{ij} (s_i, s_j) = \tanh J_{ij}.
\end{align}
For small values of the interaction terms $\{J_{ij}\}$,
the Hessian of $\mathcal{F}_\textrm{Bethe}$ at that point makes sure
that the solution is \textit{minimum} \cite{Mooij2005,Mooij2005a}.

For simplicity, 
we here assume that the interactions $\{ J_{ij} \}$ are independent Gaussian random variables
with mean $0$ and variance $1$. 
In this case,
it has been shown that at the critical inverse temperature $\beta_\textrm{c}$,
the Ising spin system causes the transition from the paramagnetic phase to a spin-glass phase
when the inverse temperature $\beta$ increases \cite{Mooij2005,Mooij2005a}.

\section{STATISTICAL-MECHANICAL ITERATIVE ALGORITHM}

\subsection{Update procedure}

\begin{figure}
\begin{center}
  \includegraphics[width=8cm,keepaspectratio,clip]{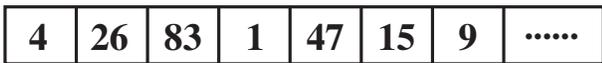} 
\caption{
One of examples of the update list.
We use node numbers in the update list in turn.
}
\label{fig_list}
\end{center}
\end{figure}

The statistical-mechanical iterative algorithm,
so-called \textit{Belief Propagation}, is often used in various problems
\cite{Pearl1988,Yedidia2001,Tanaka2002,Tanaka2003,Yedidia2005}.
The iterative algorithm calculates the solution which minimizes the Bethe free energy.
Though the statistical-mechanical iterative algorithm is often performed by a synchronous update method,
we use an asynchronous update method
to investigate effects of heterogeneous structures on the iterative algorithm.
The asynchronous update method of the statistical-mechanical iterative algorithm needs an update list
in which the node numbers to be updated are listed.
Figure~\ref{fig_list} shows one of the examples of the update list.
In this case, at first we update the information related to node $4$,
and at the next step, update the information related to node $26$.
If we reach the end of the update list, one iteration procedure is finished.

The iterative algorithm by means of the Bethe approximation is as follows:
\begin{enumerate}
\item[\textit{Step 1.}]
We set initial values of $\left \{ m_{ij}(s_j)  | i,j \in \Omega \right\}$.
These initial values are chosen randomly with an uniform distribution ($0 \leq m_{ij}(s_j) \leq 1$).
The value $m_{ij}(s_j)$ is called a message from node $i$ to node $j$ in state $s_j$.

\item[\textit{Step 2.}]
We set $\tilde{m}_{ij}(s_j) \Leftarrow m_{ij}(s_j)$.
The values $\{ \tilde{m}_{ij}(s_j)\}$ represent old states,
which are needed to evaluate a convergence condition introduced later.

\item[\textit{Step 3.}]
We set a node number in the update list as the value of $i$.
If it is the first step in one iteration procedure,
we use the first node number in the update list,
otherwise we use the next node number following the one used in the previous step.

\item[\textit{Step 4.}]
For the nearest neighbor nodes of node $i$, i.e., $j \in \partial i$,
we update $m_{ij}(s_j)$ as follows:
\begin{align}
&m_{ij}(s_j) 
\Leftarrow 
\frac{ \sum_{s_i} \phi_{ij}(s_i,s_j) \prod_{k \in \partial i \setminus j} 
m_{ki}(s_i) }
{\sum_{s_{j}'} \sum_{s_i} \phi_{ij}(s_i,s_{j}') \prod_{k \in \partial i \setminus j} m_{ki}(s_i)}.
\label{eq_update}
\end{align}

\item[\textit{Step 5.}]
Until we reach the end of the update list,
we repeat \textit{Step 3} and \textit{Step 4}.

\item[\textit{Step 6.}]
When we reach the end of the update list,
we say that one iteration procedure is finished.
We here calculate the following convergence condition:
\begin{align}
\frac{1}{2\lvert E \rvert}
\sum_{i \in \Omega} \sum_{j \in \partial i} \sum_{s_j} 
\left\lvert  
\tilde{m}_{ij}(s_j) - m_{ij}(s_j) 
\right\rvert < \epsilon.
\label{eq_convergent_condition}
\end{align}
If not convergent, we back to \textit{Step 2},
otherwise calculate $p_i(s_i) \, (i \in \Omega)$
using the following equation:
\begin{align}
p_i(s_i) = \frac{ \prod_{k \in \partial i} m_{ki}(s_i) }
{\sum_{s_{i}'} \prod_{k \in \partial i} m_{ki}(s_{i}')}.
\end{align}
\end{enumerate}
Then, a local magnetization of Ising spin $i$, $\langle s_i \rangle$, is calculated by
\begin{align}
\langle s_i \rangle \simeq \sum_{s_i = \pm 1} s_i p_i(s_i).
\end{align}
The pair probability distribution $p_{ij}(s_i,s_j)$ is calculated by the similar procedure;
the details of the iterative algorithm is denoted in Ref.~\cite{Yedidia2001,Tanaka2002,Tanaka2003}.
After the iterations, we get the solution $\{ p_i(s_i) \}$ and $\{ p_{ij}(s_i,s_j) \}$
which minimizes the Bethe free energy of Eq.~\eqref{eq_bethe}.

\subsection{Making the update list}
\begin{figure}
\begin{center}
  \includegraphics[width=8cm,keepaspectratio,clip]{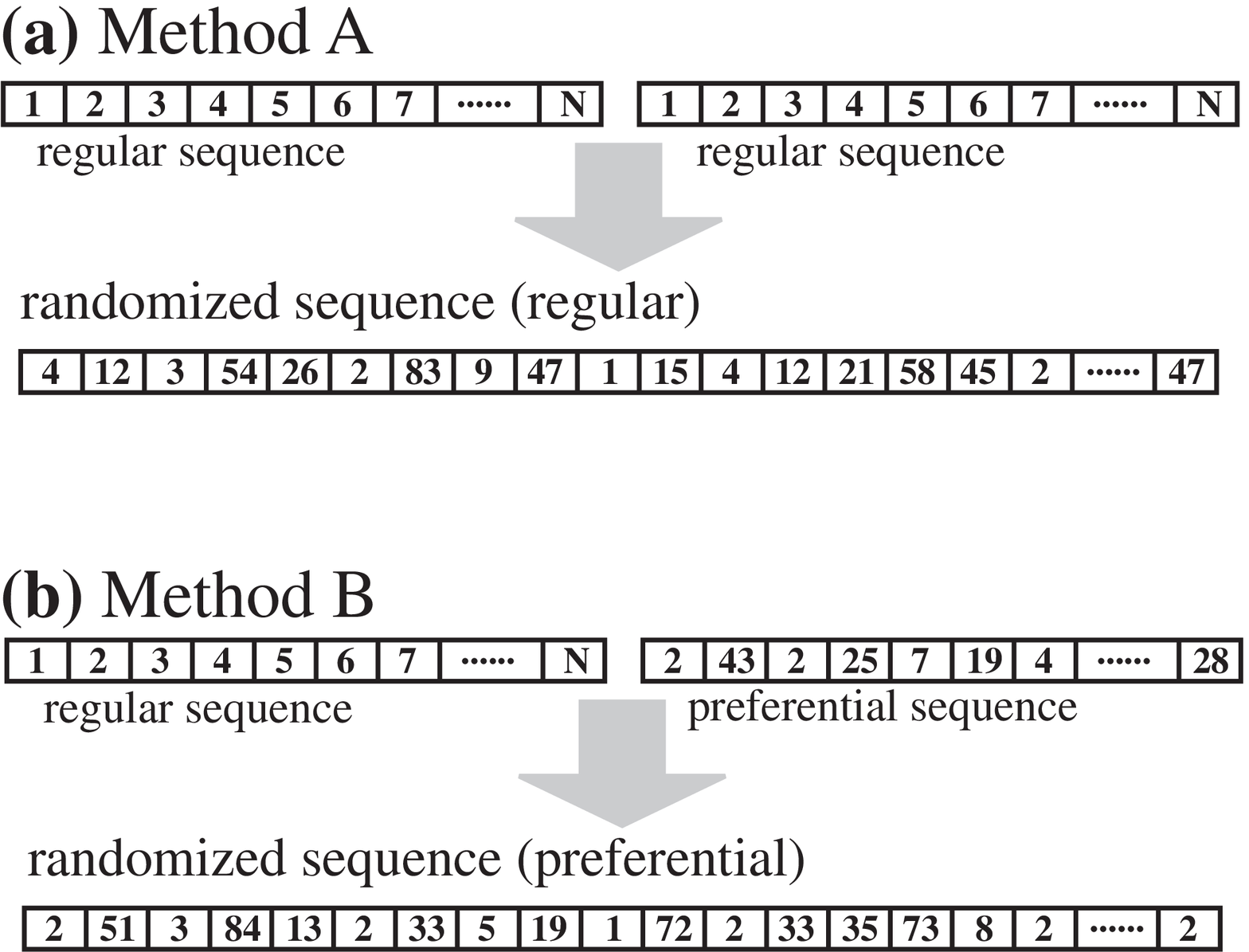} 
\caption{
Two methods for making the update list.
(a) Ordinary update list in which all node numbers are selected at random only two times.
(b) Preferential update list in which high-degree nodes are likely to chosen as update nodes.
This preferential update procedure uses the information of heterogeneity of the complex network structures.
}
\label{fig_method}
\end{center}
\end{figure}

\begin{figure}
\begin{center}
  \includegraphics[width=5cm,keepaspectratio,clip]{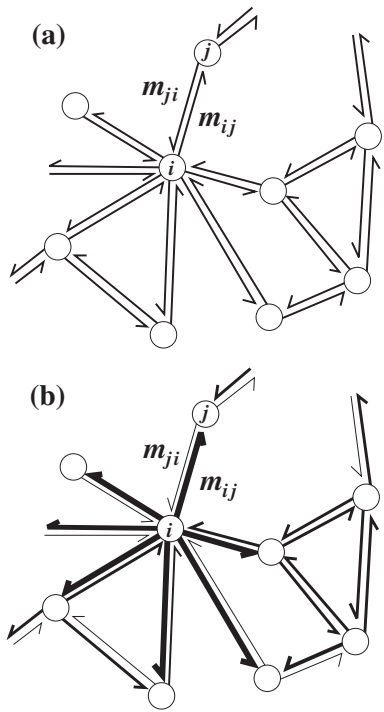} 
\caption{
Intuitive pictures of the update procedures.
(a) Ordinary update method (method A in Fig.~\ref{fig_method}.)
(b) Preferential update method (method B in Fig.~\ref{fig_method}.)
The messages represented by bolder arrows tend to be updated more frequently than those with thinner arrows.
}
\label{fig_update}
\end{center}
\end{figure}

Methods of making the update list used in the update procedure are arbitrary;
an uniformly random update list is usually used,
in which each node number is selected at random.
However, we expect the following assumption:
``\textit{Usage of heterogeneity of complex networks makes algorithms and dynamics more efficient}.''
As the simplest case,
we here assume that high-degree nodes play an important role in the scale-free networks;
the update of the information of the high-degree nodes causes large effects on the next update step
because the information of the high-degree nodes could be used a lot in the next update step.
Therefore, to investigate how large the high-degree nodes influence on the iterative algorithm,
we compare results obtained by two different update lists;
one is made by the ordinary random procedure,
and the other is generated by using the information of heterogeneity of the network structures.

These update lists are made by the following procedure.
The making procedures are illustrated in Fig.~\ref{fig_method}.
\begin{enumerate}
\item[]
\textit{Method A (ordinary update list)}
\item[(A1)]
Two regular sequences are set.
Each regular sequence has $N$ components,
and each node number is listed only one time in each sequence.
\item[(A2)]
We randomly select a number from these two sequences.
The selected number is stocked into the update list.
Once a number is selected, the number is removed from the sequences.
\item[(A3)]
We repeat the procedure (A2) until there are no number in the two regular sequences.
Note that when the procedure is finished,
there are $2N$ numbers in the update list,
and each node number is listed only two times.
\end{enumerate}

Different from method A,
we make the preferentially selected update list using the information of the network structures in method B.
In this case, we use the information of a degree of each node.
\begin{enumerate}
\item[]
\textit{Method B (preferential update list)}
\item[(B1)]
One regular sequence is set.
The regular sequence has $N$ components and each node number is listed only one time.
Additionally, one preferential sequence is set.
In the preferential sequence, each listed number is selected 
with the probability proportional to its corresponding node's degree;
hence, nodes with higher-degree tend to be selected more frequently than those with lower-degree.
\item[(B2)]
We select a number from these two sequences at random.
The selected number is stocked into the update list.
Once a number is selected, the number is removed from the sequences.
\item[(B3)]
We repeat the procedure (B2) until there are no number in the two regular sequences.
Note that when the procedure is finished, there are $2N$ numbers in the update list.
In the update list, 
each number corresponding to each node is listed at least one time,
and several numbers emerge a lot because of their high-degree.
\end{enumerate}

Intuitive pictures of effects of those methods are illustrated in Fig.~\ref{fig_update},
in which the messages denoted by bolder arrows are updated more frequently than thinner ones.
Figure~\ref{fig_update}(a) shows the update procedure using method A.
In Fig.~\ref{fig_update}(a), all arrows have the same width,
which represent that we update all messages in the same frequency.
On the other hand, in Fig.~\ref{fig_update}(b)
each arrow has a different thickness depending on its degree.
For instance, node $i$ has a lot of connections to other nodes.
In the update procedure of messages related to node $i$,
messages of many other nodes are needed, see Eq.~\eqref{eq_update}.
Furthermore, the updated messages on the node $i$ are used for the other update procedures on the many other nodes.
Since the messages on high-degree node are frequently updated in method B,
it is seems that information of local update is easily propagated to the whole system.
Then, we expect that method B works more efficiently than method A
for the heterogeneous networks like the scale-free networks.

\section{NUMERICAL RESULTS}

\subsection{The number of iteration steps}

To investigate effects of heterogeneous structures on the iterative algorithm,
we performed numerical experiments.
In the numerical experiments, two network structures were used.
One is the random network proposed by Erd{\"o}s and R{\'e}nyi (ER networks)\cite{Erdos1960},
and the other is the scale-free network proposed by Barab{\'a}si and Albert (BA networks)\cite{Barabasi1999a}.
The ER networks have homogeneous network structures,
in which each node has the similar degree though there are some fluctuations;
the degree distribution of the ER networks is the Poisson distribution.
On the other hand, the BA network is the representative of the complex networks with heterogeneity,
in which there are a few high-degree nodes, which is often called \textit{hubs},
and a lot of low-degree ones; the degree distribution is $P(k) \sim k^{-3}$.

In all the numerical experiments, we use the networks with the network size $N = 400$,
the average degree $\langle k \rangle = 6.0$.
As for the convergence condition of Eq.~\eqref{eq_convergent_condition}, we here set $\epsilon = 10^{-6}$.
If 1000 iterative procedures do not make the system convergent, 
we consider the system is not convergent and finish the update procedure.
In addition, as far as we checked, messages $\{m_{ij}\}$ are oscillated in the nonconvergent realizations.
We have performed the calculations for $200$ different realizations
in which the network structures and interaction terms $\{J_{ij}\}$ are different from each other.
All quantities discussed below are averaged over only convergent realizations.

\begin{figure}
\begin{center}
  \includegraphics[width=7cm,keepaspectratio,clip]{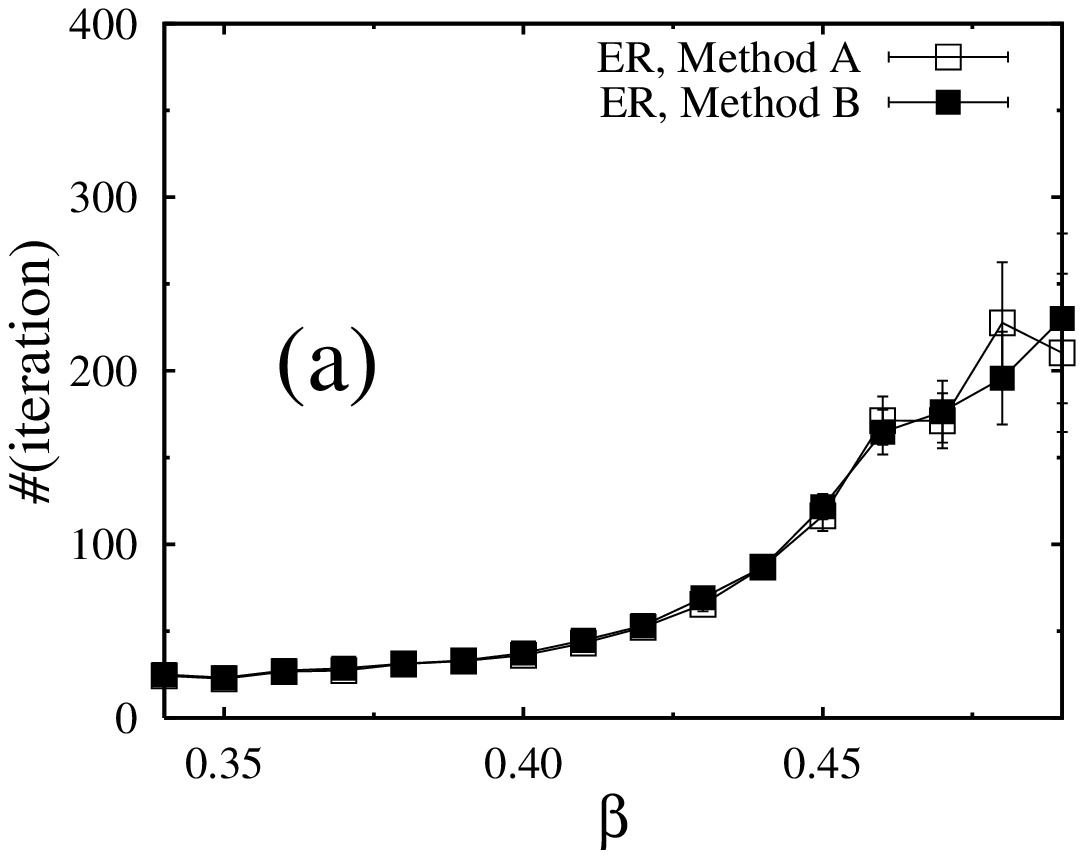} \\
  \includegraphics[width=7cm,keepaspectratio,clip]{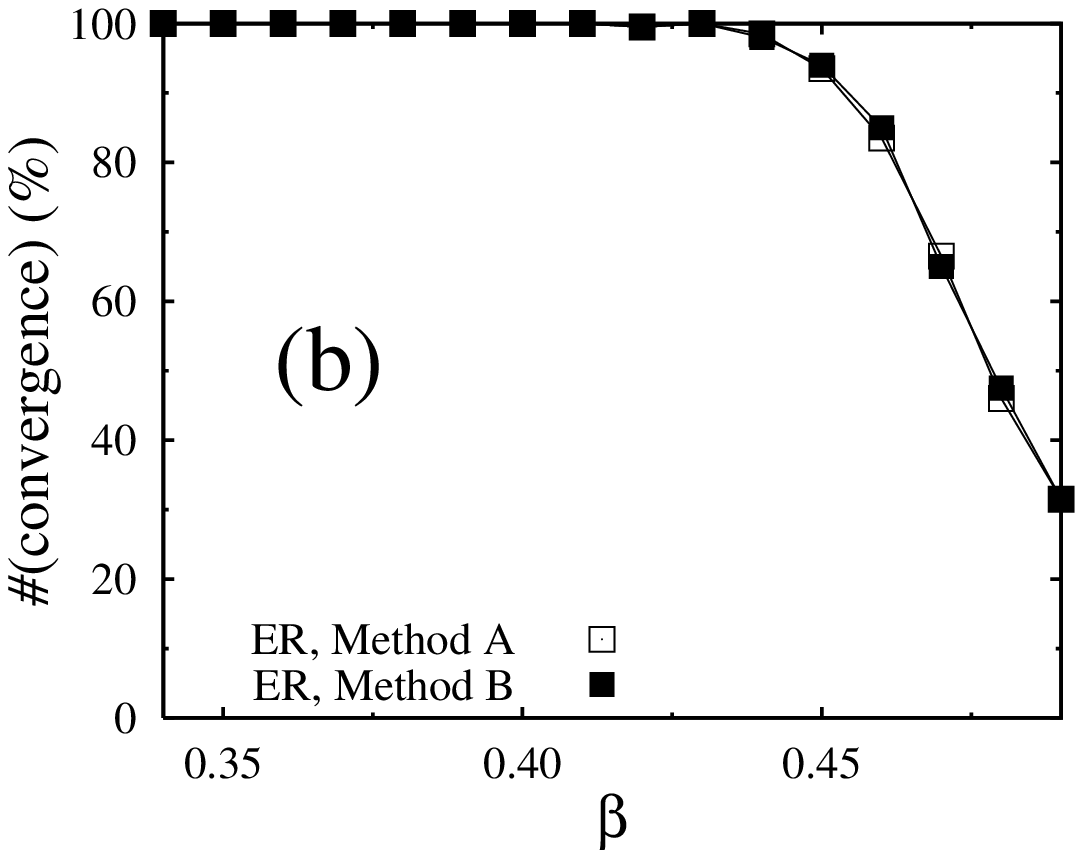} \\
  \includegraphics[width=7cm,keepaspectratio,clip]{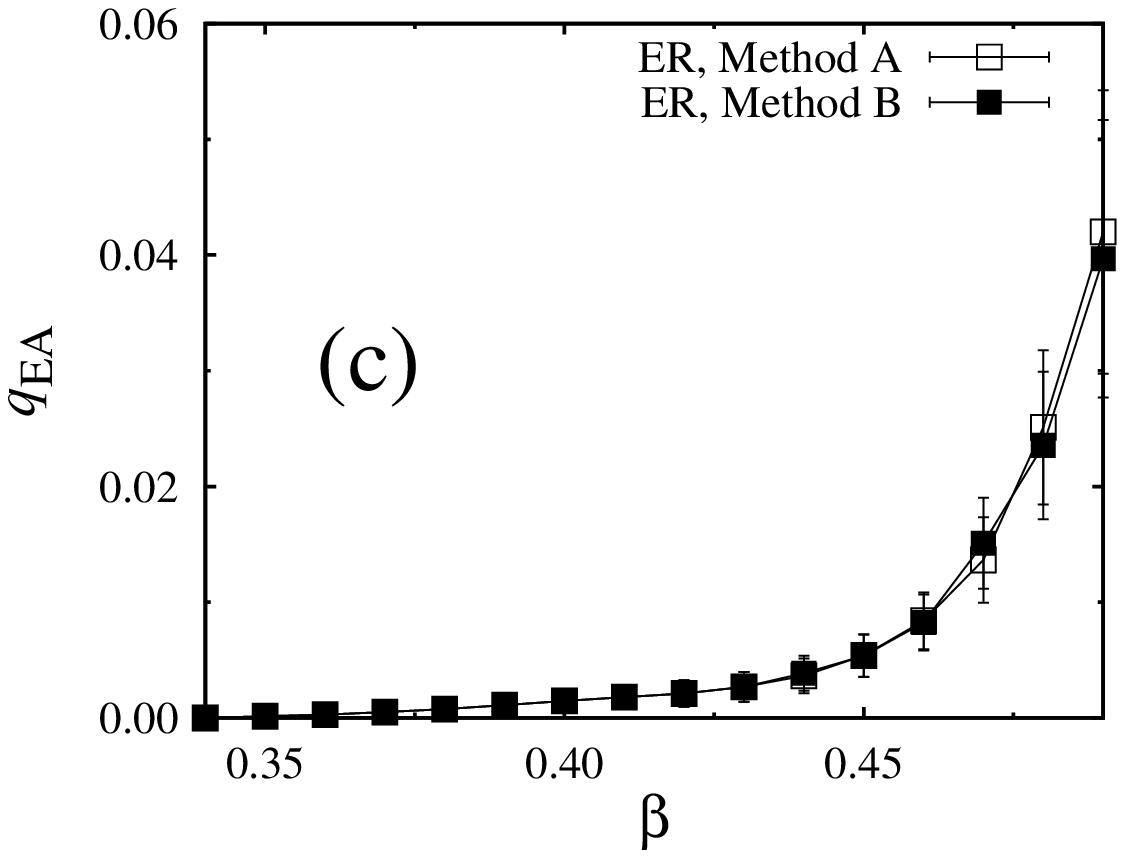} 
\caption{
Results of the numerical experiments in the case of the ER networks.
(a)The number of iteration steps needed for convergence as a function of the inverse temperature $\beta$.
(b)The percentage of convergent realizations in $200$ different realizations.
(c)The Edward-Anderson order parameter $q_\textrm{EA}$.
}
\label{fig_result_er}
\end{center}
\end{figure}

\begin{figure}
\begin{center}
  \includegraphics[width=7cm,keepaspectratio,clip]{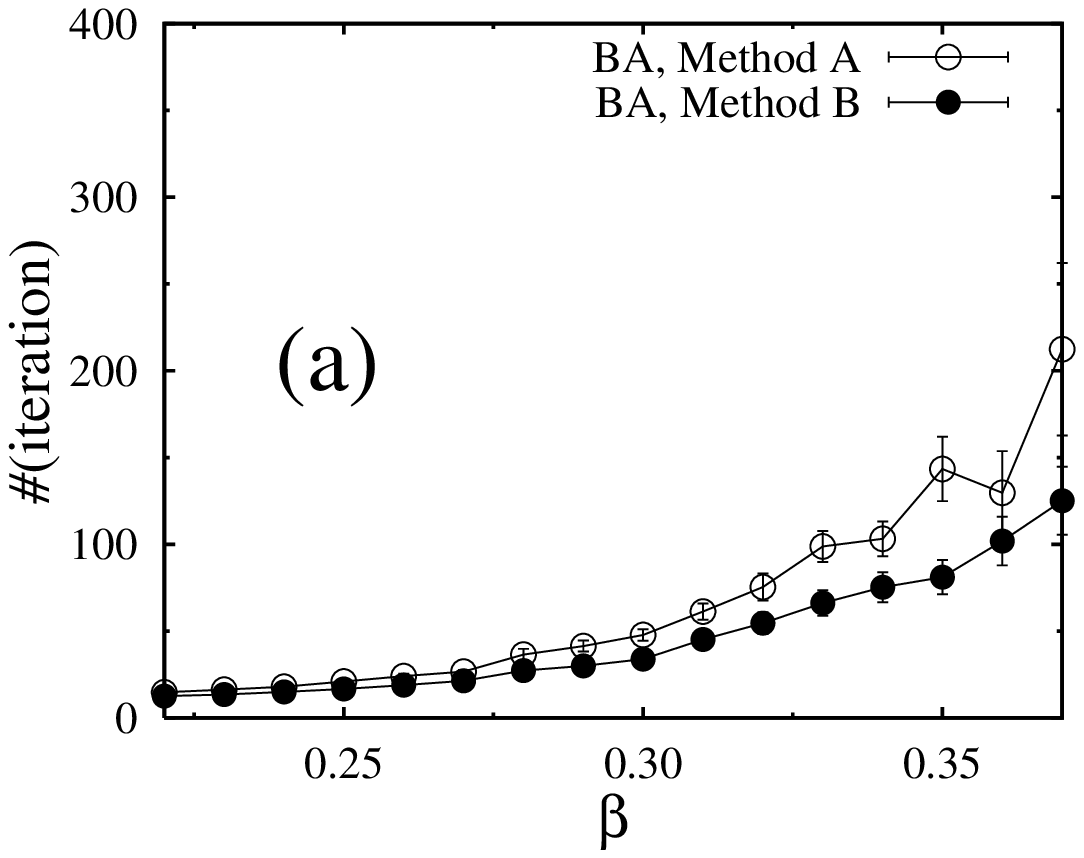} \\
  \includegraphics[width=7cm,keepaspectratio,clip]{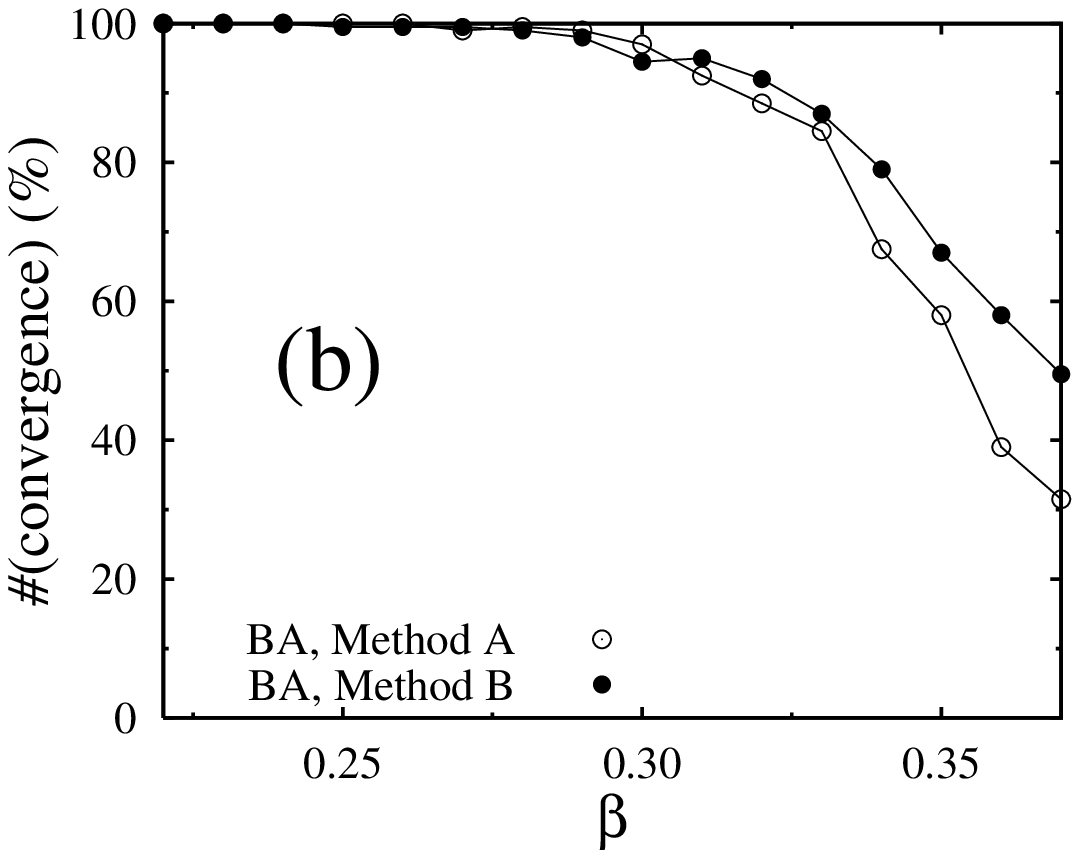} \\
  \includegraphics[width=7cm,keepaspectratio,clip]{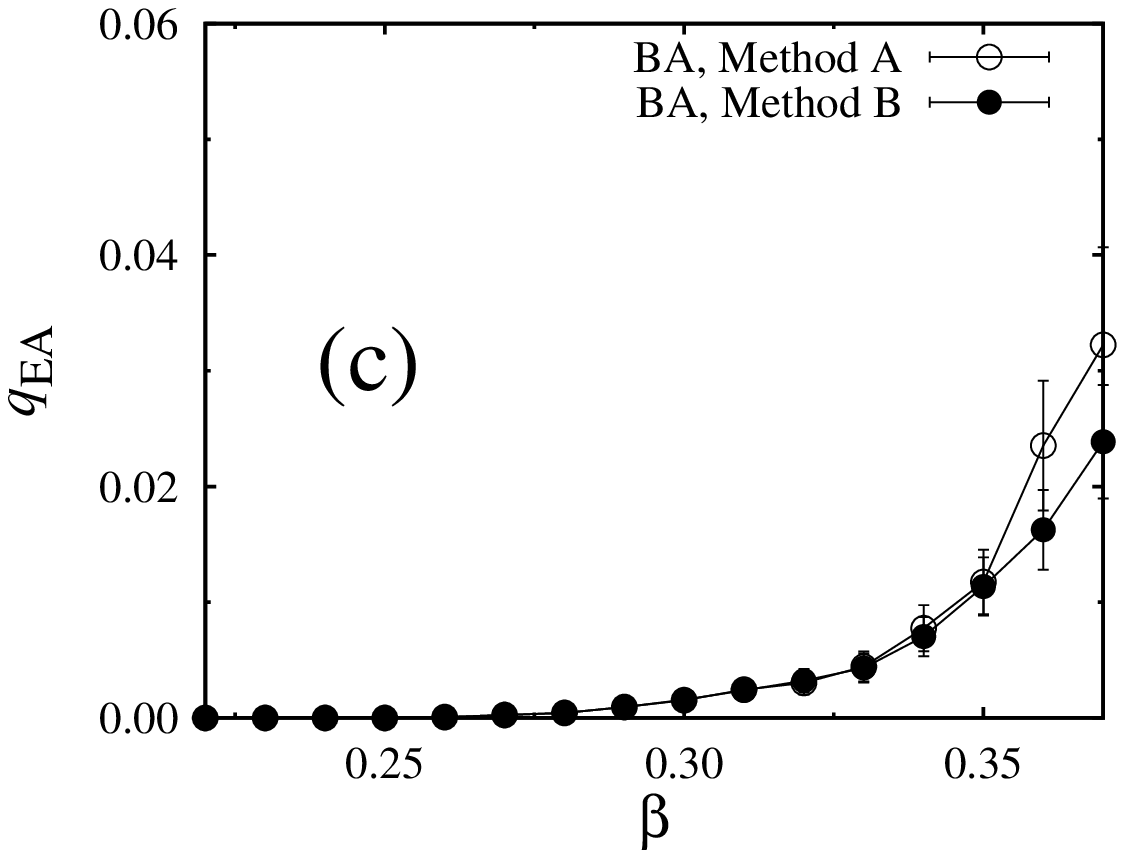} 
\caption{
Numerical results for the case of the BA networks.
(a)The number of iteration steps needed for convergence.
(b)The percentage of convergent realizations in $200$ different realizations.
(c)The Edward-Anderson order parameter $q_\textrm{EA}$.
}
\label{fig_result_ba}
\end{center}
\end{figure}

Results for the case of the ER networks are shown in Fig.~\ref{fig_result_er},
and Fig.~\ref{fig_result_ba} shows those of the BA networks. 
Figures~\ref{fig_result_er}(a) and \ref{fig_result_ba}(a) show the number of iteration procedures
needed for satisfying the convergence condition of Eq.~\eqref{eq_convergent_condition}.
The percentage of convergent realizations in $200$ different ones is shown 
in Figs.~\ref{fig_result_er}(b) and \ref{fig_result_ba}(b).
As the inverse temperature $\beta$ increases,
the number of convergent realizations decreases at some critical inverse temperatures $\beta_\textrm{c}$
in both cases of ER and BA networks,
though ER and BA networks have the different critical inverse temperatures.
We can consider the phenomenon as the phase transition from the \textit{paramagnetic phase}
to the \textit{spin-glass phase}, as discussed by Mooij and Kappen \cite{Mooij2005,Mooij2005a}.
To confirm this, we calculate the Edwards-Anderson order parameter
\begin{align}
q_\textrm{EA} \equiv \frac{1}{N} \sum_{i \in \Omega} \langle s_i \rangle^2,
\end{align} 
which becomes positive after the transition to the spin-glass phase.
Figures~\ref{fig_result_er}(c) and \ref{fig_result_ba}(c) show the Edwards-Anderson order parameter $q_\textrm{EA}$
as a function of the inverse temperature $\beta$.
At the point from which the number of convergent realizations decreases,
the Edwards-Anderson order $q_\textrm{EA}$ becomes positive \cite{footnote1};
this suggests that the system is in the spin-glass state above the critical inverse temperature $\beta$.
Note that there is not an external field
and hence the positive value of $q_\textrm{EA}$ does not indicate the ferromagnetic phase.
In the spin-glass state,
it has been shown that the statistical-mechanical iterative algorithm does not work enough well 
\cite{Mooij2005,Mooij2005a}.
While improvement of the worse convergence property of the statistical-mechanical iterative algorithm
would be needed for future works,
it is beyond the scope of the present paper.
Near the critical inverse temperature $\beta_\textrm{c}$,
we see remarkable improvement of the number of iteration steps needed for convergence
in the case of the BA networks, see Fig.~\ref{fig_result_ba}(a).
Near the critical inverse temperature $\beta_\textrm{c}$,
the number of iterative procedures needed for convergence of method B
is $0.6$ or $0.7$ times as many as that of method A.
Furthermore, the percentage of convergent realizations in the case of method B increases 
in comparison with that of method A near $\beta_\textrm{c}$.
On the contrary, for the ER random networks,
both methods A and B give similar results because of the homogeneity of the ER networks,
as shown in Figs.~\ref{fig_result_er}(a) and \ref{fig_result_er}(b).

Finally, we note the correctness of the solution. 
In the paramagnetic phase, all the numerical solutions yield $\langle s_i \rangle = 0$. 
Hence, both algorithms give the correct solution. 
However, it is hard to evaluate the correctness of the numerical solutions in the spin-glass phase; 
it is difficult to obtain the exact solution in the spin-glass phase. 
We have compared the Bethe free energies of the numerical solutions 
because a solution with a lower value of the free energy corresponds to more correct one,
and confirmed that the free energies calculated by both algorithms have similar behavior to each other.
Therefore, we consider that both algorithms give similar solutions 
while more detailed discussion would be needed in future research.

\subsection{Effect on messages}

\begin{figure}
\begin{center}
  \includegraphics[width=7cm,keepaspectratio,clip]{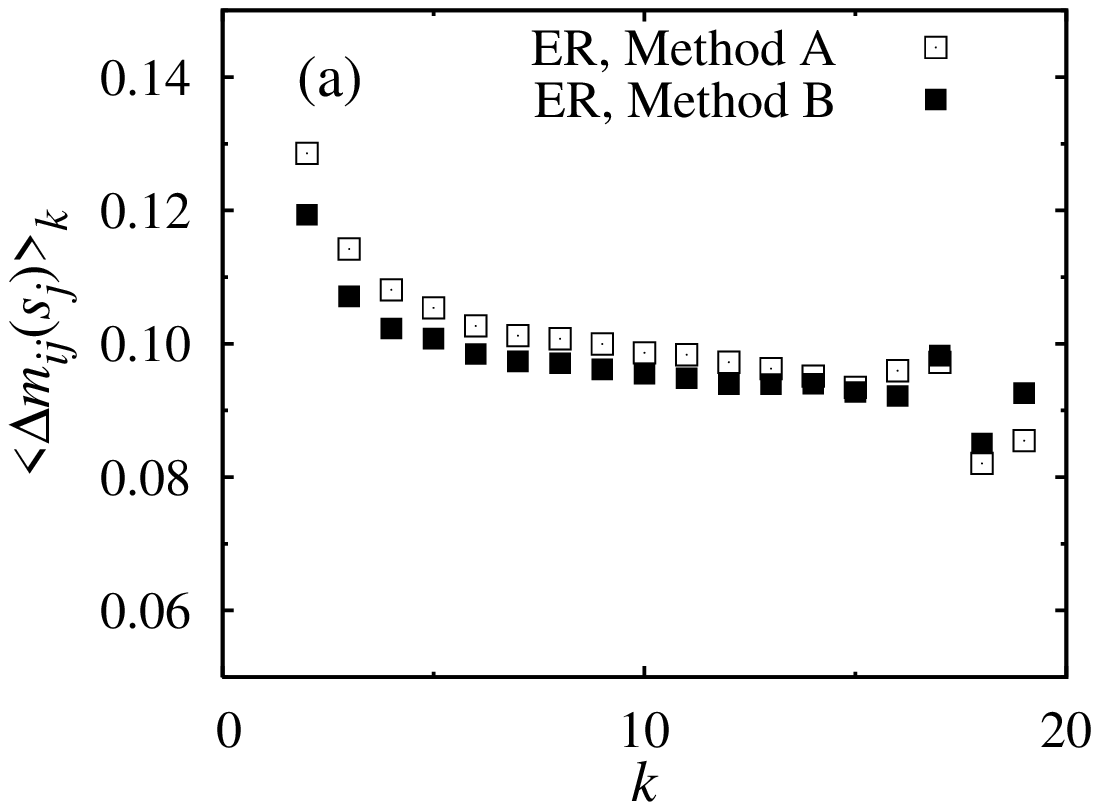} 
  \includegraphics[width=7cm,keepaspectratio,clip]{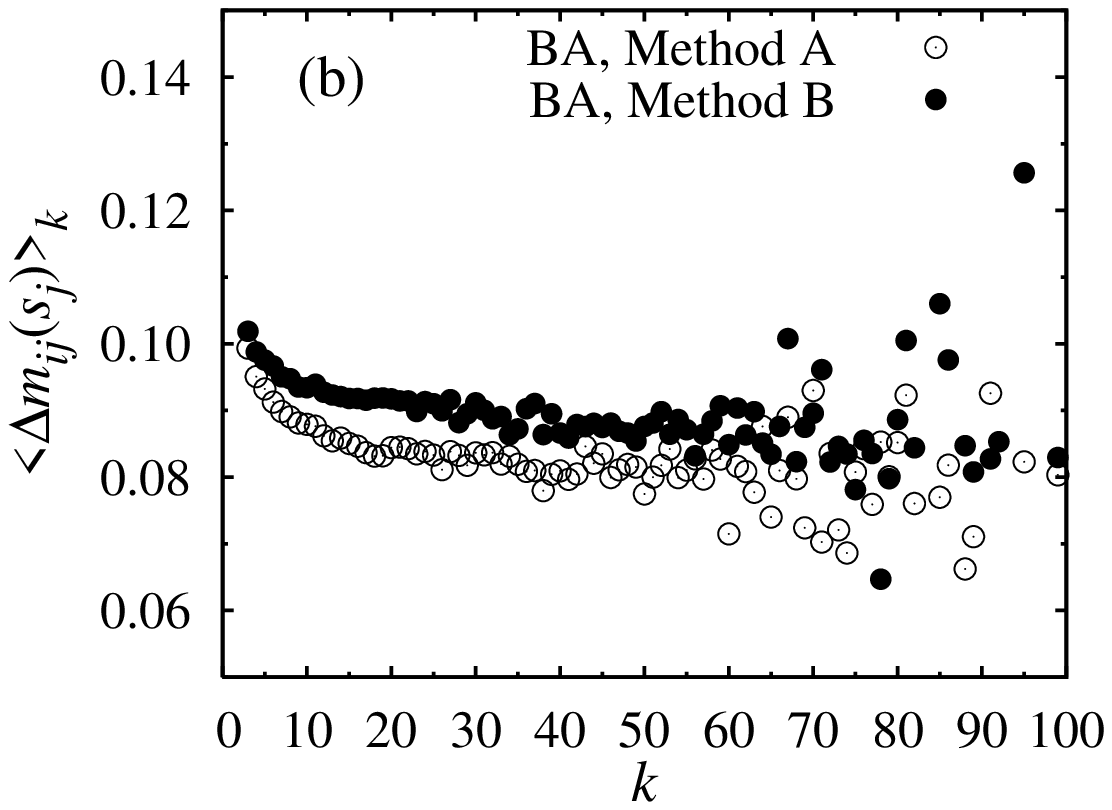} 
\caption{
The averaged changes of messages over nodes with degree $k$.
(a) in the case of ER networks.
(b) in the case of BA networks.
In the both cases, we set the inverse temperature $\beta = 0.2$.
}
\label{fig_mes}
\end{center}
\end{figure}

To investigate how the proposed update algorithm effects on the update procedure,
we next investigate the property of message $m_{ij}(s_j)$ in Eq.~\eqref{eq_update}.
We here define the following quantities to characterize the influence on the update procedure:
\begin{align}
\Delta m_{ij}(s_j) &\equiv \lvert \tilde{m}_{ij}(s_j) - m_{ij}(s_j) \rvert, \\
\langle \Delta m_{ij}(s_j) \rangle_k 
&= \left( \sum_{i \in \Omega} \delta_{k_i,k} \right)^{-1}
\sum_{i \in \Omega}
\frac{\delta_{k_i,k}}{2 \lvert \partial i \rvert} \sum_{j \in \partial i} \sum_{s_j} \Delta m_{ij}(s_j),
\end{align}
where $\delta_{k_i,k}$ is the Kronecker delta.
The quantity $\Delta m_{ij}$ shows the change of each message between the old iterative procedure 
and new iterative one.
The quantity $\langle \Delta m_{ij}(s_j) \rangle_k$
shows that the averaged change of messages over nodes with degree $k$.
Figure~\ref{fig_mes} shows the changes of the messages related to the nodes with degree $k$, 
i.e., $\langle \Delta m_{ij}(s_j) \rangle_k$.
In these numerical experiments, we used the inverse temperature $\beta = 0.2$
in which the both systems on the ER and the BA networks converged in all realizations.
In the case of the ER networks,
methods A and B have similar results as for $\langle \Delta m_{ij}(s_j) \rangle_k$,
though method A makes the change of messages a little larger, as shown in Fig.~\ref{fig_mes}(a).
On the contrary, in the case of the BA networks,
method B makes the changes of messages $\langle \Delta m_{ij}(s_j) \rangle_k$ 
larger than method A for low degree region $k \leq 60$.
This suggests that 
the frequent updates related to high-degree nodes make the change of the messages on low-degree nodes larger,
and hence, the number of iteration steps decreases 
because the larger changes of messages cause rapid convergence to the solution.

\begin{figure}
\begin{center}
  \includegraphics[width=7cm,keepaspectratio,clip]{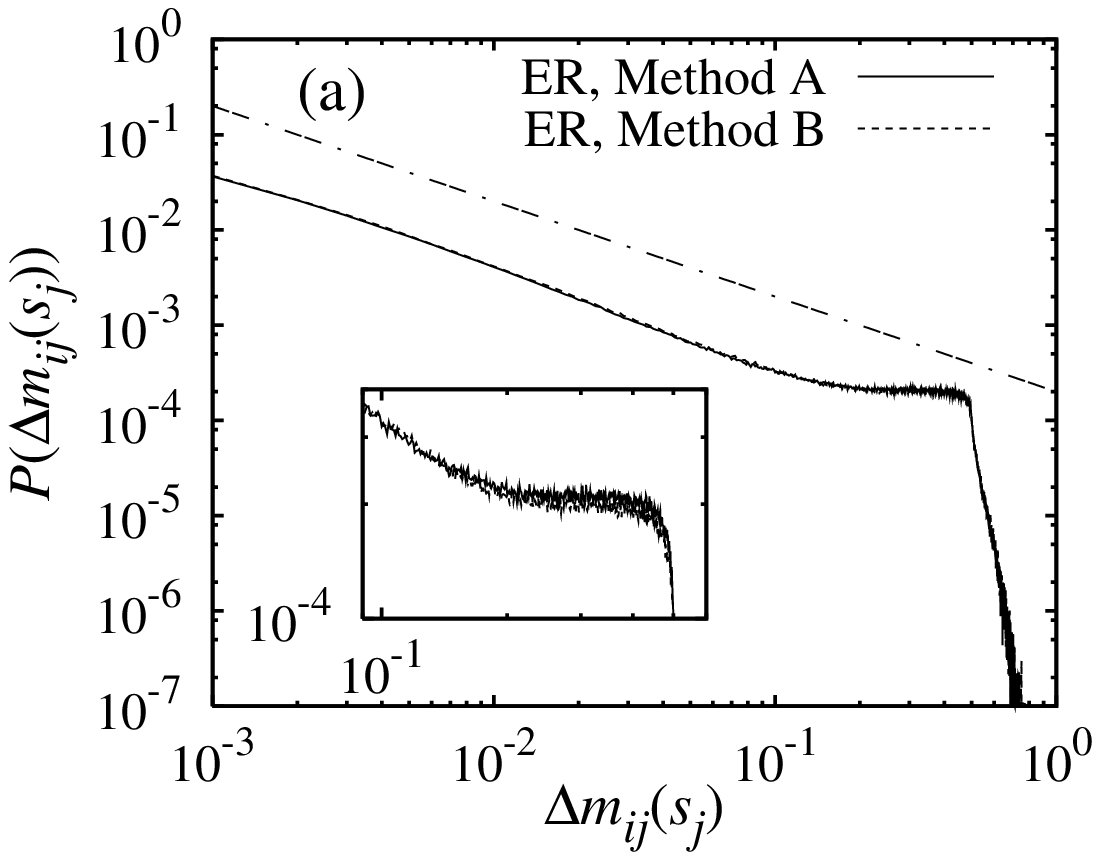} 
  \includegraphics[width=7cm,keepaspectratio,clip]{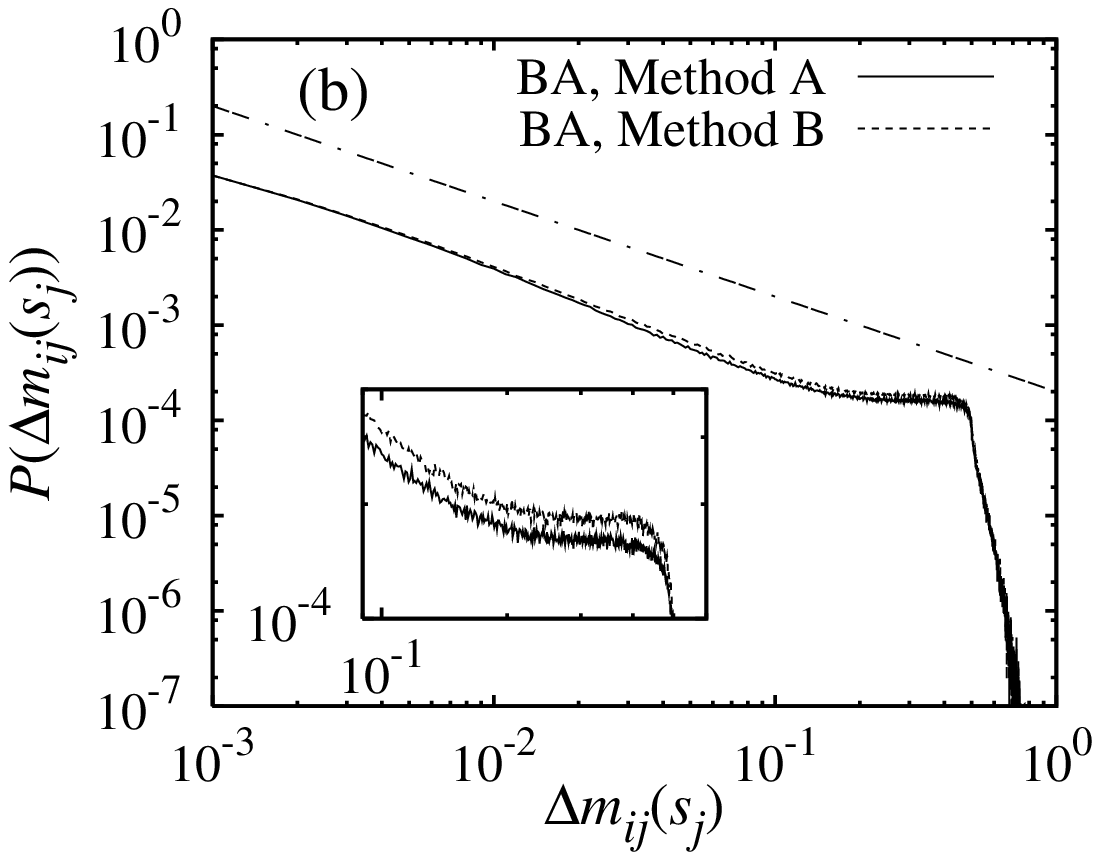} 
\caption{
Distributions of $\Delta m_{ij}(s_j)$ at $\beta = 0.2$.
(a) and (b) show the cases of the ER networks and the BA networks, respectively.
Each inset in (a) and (b) is an enlarged figure at large $\Delta m_{ij}(s_j)$ region.
The dot-dashed lines are guides for the eyes
and represent $P(\Delta m_{ij}(s_j)) \sim \Delta m_{ij}(s_j)^{-1}$.
}
\label{fig_mes_dis}
\end{center}
\end{figure}

Distributions of the change of messages $\Delta m_{ij}(s_j)$ are shown in Fig.~\ref{fig_mes_dis}(a) and \ref{fig_mes_dis}(b).
Each inset in Fig.~\ref{fig_mes_dis} is an enlarged figure at large $\Delta m_{ij}(s_j)$ region.
In the ER random networks, method B makes no difference from the case of method A.
In the BA networks, as suggested in Fig.~\ref{fig_mes}(b),
method B makes the changes of messages larger than those of method A.
The inset of Fig.~\ref{fig_mes_dis}(b) shows that method B makes the distribution of $\Delta m_{ij}(s_j)$
a little up compared to the one of method A,
which suggests that method B work well in the case of the BA networks.
It is valuable to note that in both cases
the distribution of $\Delta m_{ij}(s_j)$ shows
the power-law form with the degree exponent $-1$, as shown in Figs.~\ref{fig_mes_dis}(a) and (b),
while it is beyond the present paper to clarify why the distribution obeys a power-law form,
so-called Zipf's law.

\section{CONCLUSION}
In the present paper, 
we have investigated effects of heterogeneity of complex networks on statistical-mechanical iterative algorithms.
Numerical experiments have shown that
the usage of information of heterogeneity affects the algorithm in BA networks,
but does not influence that in ER networks.
It has been revealed that high-degree nodes play important roles in the iterative algorithms;
the usage of information of degrees
makes the statistical-mechanical iterative algorithm more efficient in the case of the BA networks.
At the critical inverse temperature, the system has transition from the paramagnetic phase to the spin-glass phase.
Near the critical inverse temperature $\beta_\textrm{c}$,
the number of iteration steps needed for convergence decrease, 
and the percentage of convergence increases in the case of the BA networks.
In the case of the ER networks, however,
the usage of information of degrees does not make differences.

The changes of messages used in the iterative algorithm have also been investigated,
and it becomes clear that 
the newly proposed iterative algorithm causes larger changes of messages in the BA networks.
This result agrees with intuitive pictures
that the system converges fast when there are the larger changes of the messages.
These results show
that the information of the whole system propagates rapidly through high-degree nodes,
and hence, the usage of the heterogeneity makes the iterative algorithm more efficient.

We conclude that the heterogeneity is important for the iterative algorithms
and the usage of the heterogeneity makes it possible to perform the iterative calculation efficiently.
While the current study is meaningful
as a first step of improvement of the statistical-mechanical iterative algorithm,
one can consider other approaches to improve the iterative algorithm.
We used the heterogeneity of degrees of nodes in the present paper,
and other heterogeneous property
such as the load or the betweenness centrality \cite{Goh2001},
community structure \cite{Newman2004},
and network motifs \cite{Milo2002}
may be available for improvement of the iterative algorithm.
We think that the improvement of the iterative algorithm is important for large complex networks.
In order to achieve the improvement, 
it could be helpful to research the usage of the heterogeneity of network structures.
Additionally, these studies would also be valuable to understand dynamical properties of complex networks.

\begin{acknowledgments}
This work was supported in part by grant-in-aid for scientific research (No. 14084203 and No. 17500134)
from the Ministry of Education, Culture, Sports, Science and Technology, Japan.
\end{acknowledgments}


\begin{thebibliography}{31}
\bibitem{footnote1}
When the system enters the spin-glass phase, the value of $\langle s_i \rangle$ at each node $i$ 
also starts to be non-zero.
\end{thebibliography}
\end{document}